# Capítulo 1

# El debate entre Ernst Mayr y Carl Sagan acerca de la probabilidad de vida inteligente en el universo[1]

**Resumen:** Durante la *Segunda Escuela Iberoamericana de Posgrado en Astrobiología,* se suscitaron interesantes debates acerca de la probabilidad de existencia de seres inteligentes extraterrestres en el universo, entre los expertos del área biológica y del área física. Por esta razón, resulta apropiado reproducir por primera vez en español un debate sobre el mismo tema desarrollado entre Ernst Mayr y Carl Sagan en el año 1995. Dicho debate había sido organizado por Guillermo A. Lemarchand y publicado en las páginas de dos números consecutivos de *Bioastronomy News*. Aquí se reproduce el debate completo, incluyendo la introducción original.

**Abstract:** During the Second Iberoamerican Graduate School on Astrobiology interesting debates, between the experts from the biological and physical backgrounds, arose about the probability of the existence of extraterrestrial intelligent beings in the universe. For this reason, it is appropriate to reproduce –for the first time in Spanish- the debate on the subject conducted, in 1995, between Carl Sagan and Ernst Mayr. This debate was organized by Guillermo A. Lemarchand and published in the pages of two consecutive numbers of *Bioastronomy News*. Here we reproduce the complete debate, including its original introduction.

---

1 Este debate fue organizado y editado originalmente en inglés por Guillermo A. Lemarchand (GAL) en el año 1995. Apareció por primera vez en dos números sucesivos de *Bioastronomy News: Newsletter of the International Astronomical Union Commission 51*, vol. 7 (3) y (4), 1995, publicado en EEUU por The Planetary Society (www.planetary.org). En 1996 el debate también el debate fue publicado nuevamente reproducido en forma completa en *The Planetary Report*, vol. XVI (3): 4-13, May/June 1996. Por el tipo de discusiones que tuvieron lugar entre profesores y estudiantes, a lo largo de la *Segunda Escuela Iberoamericana de Astrobiología*, se consideró muy apropiado reproducir este debate entre dos figuras emblemáticas de la ciencia del siglo XX como Ernest Mayr y Carl Sagan. Los argumentos esgrimidos hace 15 años, siguen representando muy bien las diferentes posiciones sobre la posibilidad de vida inteligente en el universo que asumen los representantes de las ciencias biológicas y físicas respectivamente. Esta es la primera vez que se publica el texto en la lengua castellana. Agradecemos a The Planetary Society (Pasadena) por habernos dado la autorización de traducirlo y reproducirlo en las páginas de este libro. GAL

## 1.    Introducción

Desde que los primeros seres humanos levantaron la vista hacia los cielos, comenzaron a proyectar en ellos los fantasmas de sus mentes soñadoras. Si es que existe algún hilo conductor que une a los antiguos filósofos griegos y a los modernos científicos espaciales, es la incertidumbre acerca de la pluralidad de los mundos habitados en el universo. El vasto y antiguo cosmos que se despliega ante nosotros, se escapa más allá del entendimiento humano común y nos hace reflexionar acerca del significado último de la exquisita vida que florece en nuestro delicado planeta azul.

A través del desarrollo de la tecnología y nuestro entendimiento acerca de las leyes de la naturaleza, la humanidad se encuentra por primera vez en la historia, en una posición única capaz de corroborar o refutar mediante pruebas experimentales, la hipótesis acerca de la existencia de civilizaciones tecnológicas de origen extraterrestre. El programa SETI (Search for Extraterrestrial Intelligence) que nuestra civilización humana ha comenzado a desarrollar desde principios de la década del sesenta, en el sentido más profundo, es una búsqueda acerca de nuestros orígenes, para determinar qué posición ocupamos en la historia de la vida y el universo.

Asumimos que la vida es una consecuencia natural de las leyes físicas que actúan en ambientes propicios, y esta secuencia de procesos físicos –como sucedió en la Tierra– puede ocurrir en otros lugares (Principio de Mediocridad).

Los defensores de SETI argumentan que nuestra galaxia tiene cientos de miles de millones de estrellas, y que vivimos en un universo con miles de millones de galaxias, por lo que la vida debiera ser un acontecimiento común en este ámbito cósmico. Debe haber muchos planetas habitables, cada uno de ellos refugiando a su camada de seres vivos. Algunos de estos mundos habrían de desarrollar la inteligencia, la capacidad tecnológica e interés en comunicarse con otras criaturas inteligentes. Por medio de las ondas electromagnéticas, es posible establecer contacto a través de distancias interestelares e intercambiar información y sabiduría con el resto de los vecinos cósmicos. En estos instantes, alguna de las hipotéticas civilizaciones tecnológicas de origen extraterrestre podría estar transmitiendo una determinada firma electromagnética que seríamos capaces de reconocer a través de nuestras observaciones astronómicas.

Pero debido a que aún no hemos podido encontrar una sola evidencia concreta de inteligencia extraterrestre, ha surgido una batalla filosófica entre los que podrían ser llamados *optimistas* del contacto con civilizaciones extraterrestres- que por lo general se adhieren a la visión ortodoxa de SETI - y los proponentes de la hipótesis de la singularidad, la cual sugiere que la Tierra es, probablemente, es el único planeta en el cual la vida ha desarrollado una inteligencia superior capaz de desarrollar tecnologías que permitirían la comunicación interestelar.

Aquí se presentan ambos lados del debate filosófico y científico. Primero, uno de los más prominentes especialistas en evolución del siglo XX, Ernst Mayr (1904-2005), del Museo de Zoología Comparada la Universidad de Harvard, presenta los principales argumentos de la hipótesis de unicidad. Mayr destaca que, la historia de la vida en la Tierra, presenta hechos acerca de la unicidad de la secuencia de eventos que permitió que la vida desarrolle la inteligencia. Deduce que la probabilidad de repetición de dichos eventos es despreciable lo que introduce un verdadero problema para SETI. Por otro lado, Carl Sagan (1934-1996) profesor del Departamento de Astronomía y Director del Laboratorio de Estudios Planetarios de la Universidad de Cornell, responde a los argumentos de Mayr y expresa la visión optimista.

¿Cuál de las dos posiciones será la más apetecible para su paladar? Continúe leyendo y decida por usted mismo.

Guillermo A. Lemarchand
Editor, Bioastronomy News
Junio de 1995

# 2. Es muy improbable que el proyecto SETI tenga éxito

Por Ernst Mayr

## ¿Cuál es la probabilidad de éxito del programa de búsqueda de inteligencia extraterrestre?

La respuesta a este interrogante depende de una serie de probabilidades. Hace unos años hice un análisis detallado de este problema en una publicación en alemán (Mayr 1992) y basada en ella, intentaré presentar aquí los hallazgos esenciales de dicha investigación. Mi metodología consiste en formular una serie de preguntas que analizan las probabilidades de éxito.

## ¿Cuán probable es que exista vida en otro lugar del universo?

Hasta los críticos más escépticos del proyecto SETI responderían a esta pregunta con optimismo. Las moléculas necesarias para el origen de la vida, tales como aminoácidos y ácidos nucleídos, han sido identificadas en el polvo cósmico, junto con otras macromoléculas, por lo tanto parecería bastante concebible que la vida pudiese originarse en algún otro lugar del universo.

Por otro lado, algunos de los modernos escenarios sobre el origen de la vida proponen un inicio a partir de moléculas aún más simples –un comienzo que hace que un origen de la vida independiente y múltiple sea un escenario aún más probable. Sin embargo, un escenario de origen de la vida independiente y múltiple, presumiblemente, resultaría en seres vivientes drásticamente diferentes a los de la Tierra.

## ¿En qué lugar podría esperarse encontrar vida como tal?

Obviamente, sólo en planetas. A pesar de que hasta hoy tenemos conocimiento certero sólo de los nueve planetas de nuestro sistema solar[2], no hay razón alguna para dudar de que en todas las galaxias deba haber millones sino miles de millones de planetas. El número exacto, por ejemplo, para nuestra propia galaxia, sólo podríamos intentar adivinarlo.

---

2 NOTA DEL EDITOR: en el momento en que se escribió este artículo solo se habían descubierto planetas alrededor de "pulsares". La primera evidencia de planetas alrededor de estrellas tipo solar, aparecieron apenas unos meses después de la primera publicación de este artículo. En noviembre de 2010 ya se habían confirmado el descubrimiento de 497 planetas extrasolares o exoplanetas.

## ¿Cuántos de estos planetas podrían haber sido adecuados para el origen de la vida?

Evidentemente, existen grandes restricciones para la posibilidad de que la vida pueda originarse y mantenerse en un planeta. Éste debe contar con una temperatura promedio favorable; la variación estacional no debe ser muy extrema; el planeta debe estar a una distancia apropiada de su estrella central; debe tener una masa adecuada para que su gravedad sea capaz de mantener una atmósfera; la atmósfera debe tener una composición química adecuada para albergar formas de vida primitiva; debe tener la consistencia necesaria para proteger la nueva vida de la radiación ultravioleta y de otras radiaciones dañinas; y debe existir agua en dicho planeta. En otras palabras, todas las condiciones ambientales deben ser las adecuadas para propiciar el origen y el mantenimiento de la vida.

Solo uno de los nueve planetas de nuestro sistema solar tuvo la combinación adecuada de estos factores. Seguramente esto fue sólo una casualidad.

## ¿Qué fracción de planetas de otros sistemas solares tendrán una combinación igualmente adecuada de factores ambientales? ¿Será uno en 10, o uno en 100, o uno en 1.000.000?

El número que se elija dependerá del propio optimismo. Siempre resulta difícil extrapolar a partir de un solo caso conocido. Este número es, sin embargo, de suma importancia cuando se considera el limitado número de planetas que puede ser alcanzado a través de cualquiera de los proyectos de observación SETI que se desarrollan en la actualidad.

## ¿Qué porcentaje de planetas en los cuales se ha originado la vida producirán vida inteligente?

La respuesta que podrían dar los físicos, en su conjunto, a esta pregunta será completamente diferente a la que darían los biólogos. Los físicos todavía tienden a pensar más determinísticamente que los biólogos. Tienden a decir: si la vida se ha originado en algún lado, también desarrollará inteligencia a su debido tiempo. El biólogo, por otro lado, está impresionado por la improbabilidad de tal desarrollo.

La vida en la Tierra se originó alrededor de 3.800 millones de años atrás, pero la inteligencia avanzada no se desarrolló hasta hace cerca de medio millón de años atrás. Si la Tierra hubiese sido temporalmente enfriada o calentada en demasía durante esos 3.800 millones de años, la inteligencia nunca se habría desarrollado.

Al responder esta pregunta, uno debe ser consciente del hecho de que la evolución nunca se mueve en una línea recta hacia un objetivo ("inteligencia") como sucede durante un proceso químico o como resultado de una ley de la física. Las vías evolutivas son altamente complejas y se asemejan más a las bifurcaciones de las ramas de un árbol.

Después del origen de la vida, es decir, hace 3.800 millones de años, la vida en la Tierra estuvo conformada durante más de 2.000 millones de años únicamente por procariotas, simples células, sin un núcleo organizado. Estas bacterias y sus parientes desarrollaron seguramente 50 a 100 linajes diferentes (algunos quizás muy diferentes) pero, en este tiempo enormemente extenso, ninguna de ellas conllevó a la inteligencia. Debido a un único evento sorprendente que aún hoy está sólo parcialmente explicado, hace unos 1.800 millones de años, se originó la primera célula eucariota, una criatura con un núcleo bien organizado y otras características de organismos "superiores". Del rico mundo de los protistas (formados por una sola célula) se originaron eventualmente tres grupos de organismos multicelulares: hongos, plantas y animales. Pero ninguna de las millones de especies de hongos y plantas fue capaz de producir inteligencia.

Los animales (metazoos) se ramificaron en los períodos del Precámbrico y el Cámbrico en 60 a 80 linajes (filos). Sólo uno de ellos, el de los *cordados*, conllevó eventualmente a la inteligencia genuina. Los cordados son un grupo antiguo y bien diversificado, pero sólo uno de sus numerosos linajes, el de los vertebrados, produjo eventualmente la inteligencia. Entre los vertebrados, evolucionaron una serie completa de grupos – tipos de peces, anfibios, reptiles, aves y mamíferos. Nuevamente, sólo un linaje, el de los mamíferos, condujo a la inteligencia. Los mamíferos tuvieron una larga historia evolutiva que comenzó en el Período Triásico, hace más de 200 millones de años, pero fue en la última parte del Período Triásico -15 a 20 millones de años atrás- que la inteligencia superior se originó en uno de alrededor de 24 órdenes de mamíferos.

La elaboración del cerebro de los homínidos comenzó hace menos de 3 millones de años, y la de la corteza del *Homo sapiens* ocurrió hace solamente 300.000 años. Nada demuestra la improbabilidad del origen de la inteligencia superior mejor que los millones de linajes filogenéticos que no lograron conseguirla.

### ¿Cuántas especies han existido desde el origen de la vida?

Este número es motivo de especulación, tal como lo es el número de planetas en nuestra galaxia. Pero si hay 30 millones de especies vivientes, y si la expectativa de vida de una especie es de cerca de 100.000 años, entonces uno puede postular que ha habido miles de millones, quizás 50.000 millones de especies desde el origen de la vida. Sólo una de éstas adquirió el tipo de inteligencia necesaria para establecer una civilización.

Es difícil proporcionar un número exacto porque el rango de variación, tanto en el origen de las especies como en su expectativa de vida, es enorme. Las especies populosas, usualmente halladas por los paleontólogos, ampliamente difundidas y de larga duración geológica (millones de años), son probablemente excepcionales, más que típicas.

### ¿Por qué la inteligencia superior es tan poco común?

Las adaptaciones que son favorecidas por la selección, como los ojos o la bioluminiscencia, se originan en la evolución independientemente innumerables veces. La inteligencia superior se ha originado sólo una vez, en los seres humanos. Sólo se me ocurren dos posibles razones que explicarían esta singularidad. Una es que la inteligencia superior no es favorecida en absoluto por la selección natural, contrariamente a lo que esperaríamos. De hecho, todos los otros tipos de organismos vivos, millones de especies, viven bien sin inteligencia superior.

La otra razón posible para la singularidad de la inteligencia es que es extraordinariamente difícil adquirirla. Es posible detectar cierto grado de inteligencia sólo en algunos animales de sangre caliente (aves y mamíferos), lo cual no es sorprendente ya que el cerebro requiere energías extremadamente altas. Pero aún así existe un gran paso desde "cierto grado de inteligencia" a "inteligencia superior".

El linaje homínido se separó del chimpancé hace cerca de 5 millones de años, pero el gran cerebro del hombre moderno fue adquirido hace menos de 300.000 años. Como sugirió un científico (Stanley 1992), se requirió una emancipación total de la vida arbórea para hacer que los brazos de las madres estuviesen disponibles para cargar los bebés indefensos durante las etapas finales de la formación cerebral. Tan solo un 6 por ciento de las formas de vida dentro de la línea homínida, desarrollaron un cerebro grande que facilitó el surgimiento de la inteligencia superior. Parece que se requiere una compleja combinación de circunstancias improbables y favorables para producir la inteligencia superior (Mayr 1994).

### ¿Cuánta inteligencia se necesita para producir una civilización?

Como se mencionó anteriormente, es posible detectar rudimentos de inteligencia en aves (cuervos, loros) y en mamíferos no homínidos (carnívoros, delfines, monos, simios y más), pero ninguno de estos niveles de inteligencia alcanza para fundar una civilización.

### ¿Toda civilización es capaz de mandar señales al espacio y de recibirlas?

La respuesta es claramente no. En los últimos 10.000 años han habido al menos 20 civilizaciones en la Tierra, desde la del Valle del Indo, Sumeria, y otras civilizaciones del Oriente cercano, hasta Egipto, Grecia, y toda la serie de civilizaciones europeas, así como los Mayas, los Aztecas, los Incas y las tantas civilizaciones chinas e indias. Sólo una de éstas alcanzó un nivel tecnológico que le ha posibilitado enviar señales artificiales al espacio y recibirlas.

### ¿Estarán los órganos sensitivos de los seres extraterrestres adaptados para recibir nuestras señales electrónicas?

Esto no es para nada seguro. Aún en la Tierra muchos grupos de animales están especializados para estímulos olfativos o químicos, y no reaccionarían a señales electrónicas. Ni las plantas ni los hongos son capaces de recibir señales electrónicas. Aunque hubiera organismos avanzados en algún planeta, sería bastante improbable que hayan desarrollado los mismos órganos sensitivos que nosotros.

## ¿Por cuánto tiempo puede una civilización recibir señales?

Todas las civilizaciones tienen una duración limitada. Trataré de enfatizar la importancia de este punto contando una pequeña fábula.

Asumamos que existen seres realmente inteligentes en otro planeta en nuestra galaxia. Hace mil millones de años sus astrónomos descubrieron la Tierra y llegaron a la conclusión de que este planeta podría tener las condiciones propicias para producir inteligencia. Para verificar esto, mandaron señales a la Tierra durante mil millones de años sin tener una sola respuesta. Finalmente, en el año 1800 (de nuestro calendario) decidieron mandar señales durante 100 años más. Para el año 1900, no habían obtenido ninguna respuesta, por lo que concluyeron que seguramente no había vida inteligente en la Tierra.

Esto muestra que aún si hubiera miles de civilizaciones en el universo, la probabilidad de establecer una comunicación exitosa sería muy estrecha debido a la corta duración de la "ventana abierta" de comunicación.

Uno no debe olvidar que el rango de los sistemas de observación radioastronómica y óptica de SETI son muy limitados, teniendo la capacidad de detectar señales de una intensidad que sólo podrían originarse en parte de nuestra galaxia. El hecho de que haya un número casi infinito de galaxias adicionales en el universo, es irrelevante en lo que concierne a los proyectos SETI.

## Conclusiones: Una improbabilidad de dimensiones astronómicas

¿Qué conclusiones debemos sacar de estas consideraciones? No menos de seis de las ocho condiciones necesarias para el éxito de los programas de observación SETI son muy poco probables. Cuando multiplicamos estas seis probabilidades tan bajas obtenemos una improbabilidad de dimensiones astronómicas.

No obstante: ¿por qué hay tantos defensores del proyecto SETI? Cuando uno observa sus profesiones, encuentra que se trata casi exclusivamente de astrónomos, físicos e ingenieros. Simplemente no son conscientes del hecho de que el éxito de cualquier proyecto SETI no es una cuestión de leyes físicas y capacidades tecnológicas, sino que es esencialmente un tema de factores biológicos y sociológicos. Estos, obviamente, no han sido tenidos en cuenta en los cálculos del éxito posible de cualquier proyecto SETI.

## Referencias

## 3.   La Abundancia de planetas que albergan vida

Por Carl Sagan

Vivimos en una época de notable exploración y descubrimiento. La mitad de las estrellas cercanas, similares al Sol, disponen de discos circumestelares de gas y polvo, equivalentes al que tuvo la nebulosa solar de la cual se formó nuestro planeta hace 4.600 millones de años. Por medio de una técnica inesperada –residuos de los retardos temporales de las señales de radio– hemos descubierto dos planetas similares a la Tierra alrededor del pulsar B1257+12. Recientemente, también fueron detectados astrométricamente, planetas de la masa de Júpiter alrededor de las estrella 51 *Pegasi,* 70 *Virginis* y 47 *Ursae Majoris.*

Una nueva gama de técnicas utilizadas desde la Tierra o el espacio –incluyendo la astrometría, la espectrofotometría, la medición de las velocidades radiales, el uso de óptica adaptativa e interferometría- parecen estar a punto de detectar nuevos planetas jovianos, si es que existen, alrededor de las estrellas más cercanas.

Al menos una propuesta (El Proyecto FRESIP: Frecuencia de Planetas Interiores del Tamaño de la Tierra, un sistema espectrofotométrico montado en el espacio) promete detectar planetas terrestres con mayor facilidad que los jovianos. De no existir una interrupción en el financiamiento, probablemente estaremos entrando en una era dorada en el estudio de los planetas de otras estrellas de la Vía Láctea[3].

No obstante, un planeta de masa terrestre, no tiene porqué ser un planeta terrestre. Consideremos el ejemplo alternativo de Venus. Pero existen medios por los cuales, incluso desde la posición estratégica de la Tierra, podemos investigar este tema. Podemos buscar el patrón espectral del agua que se requiere para sostener océanos. Podemos buscar oxígeno y ozono en la atmósfera planetaria.

Podemos buscar moléculas como el metano, en un desequilibrio termodinámico con el oxígeno, tal que sólo pueda ser producido por la vida. De hecho, todas estas pruebas fueron realizadas exitosamente por la misión *Galileo* en sus encuentros cercanos con la Tierra en 1990 y 1992 al retomar su camino hacia Júpiter (Sagan *et al.*, 1993).

---

3   NOTA DEL EDITOR: Esta frase de Carl Sagan, escrita originalmente en 1995, se ha tornado en una realidad tangible mediante la implementación de proyectos de la envergadura del telescopio espacial KEPLER.

Las mejores estimaciones sobre el número y espaciado de planetas de masa terrestre en los sistemas planetarios en etapas tempranas de formación, como informó George Wetherill, en la primer conferencia internacional sobre zonas de habitabilidad circumestelares (Doyle, 1995), combinados con las mejores estimaciones actuales acerca de la estabilidad de océanos a largo plazo en una variedad de planetas (como informó James Kasting en la misma conferencia), sugieren que existen entre uno y dos planetas terrestres alrededor de cada estrella de tipo solar. Las estrellas mucho más masivas que el Sol, son comparativamente más escasas y envejecen rápido. Se espera que las estrellas comparativamente menos masivas que el Sol tengan planetas terrestres, pero los planetas que son suficientemente cálidos para la vida posiblemente estén anclados por fuerzas de marea, haciendo que una cara esté siempre apuntando hacia su estrella central. Sin embargo, en estos mundos, los vientos podrían llegar a redistribuir el calor de un hemisferio al otro, y su habitabilidad potencial recién está siendo motivo de estudio y análisis.

No obstante, la mayoría de la evidencia actual sugiere la existencia de un gran número de planetas distribuidos a través de la Vía Láctea con abundante agua líquida estable durante miles de millones de años. Algunos serán apropiados para la vida –nuestro tipo de vida, de carbono y agua líquida– durante miles de millones de años menos que la Tierra, algunos durante miles de millones de años más. Y, por supuesto, la Vía Láctea es una entre un sinnúmero, quizás cien mil millones, de galaxias.

## ¿La inteligencia necesariamente evolucionaría en un mundo habitado?

A partir de las estadísticas de los cráteres lunares, calibrados por muestras traídas a través de las misiones Apolo, sabemos que la Tierra sufrió, hace aproximadamente 4.000 millones de años, un bombardeo infernal de grandes y pequeños mundos provenientes desde el espacio. Este golpeteo fue lo suficientemente severo como para arrastrar atmósferas y océanos enteros hacia el espacio exterior. Antiguamente, la totalidad de la corteza Terrestre estaba constituida por un océano de magma. Claramente, este no fue un terreno fértil para la vida.

Sin embargo, poco después –Mayr adopta el número de 3.800 millones de años atrás– surgieron los primeros organismos (de acuerdo con la evidencia fósil). Presumiblemente, el origen de la vida debe haber ocurrido poco antes de esto. Tan pronto las condiciones se hicieron favorables, la vida comenzó en nuestro planeta a una velocidad sorprendente. He usado este hecho (Sagan 1974) para argumentar que el origen de la vida debe ser una circunstancia altamente probable. Ni bien las condiciones ambientales lo permiten, ¡entonces la vida aparece!

Ahora bien, reconozco que esto es a lo sumo un argumento de plausibilidad y poco más que una extrapolación a partir de un único ejemplo. Pero lo mejor que podemos hacer es restringirnos a los datos.

## ¿Es posible aplicar un análisis similar a la evolución de la inteligencia?

La Tierra es un planeta rebosante de vida, cuyo ambiente está en constante cambio, desde hace 2.000 millones de años dispone de una atmósfera rica en oxígeno, atravesó la elegante diversificación que Mayr resumió anteriormente – y en donde durante casi 4.000 millones de años no emergió nada que se parezca ni remotamente a una civilización tecnológica.

En los comienzos de estos debates (por ejemplo, Simpson 1964), los autores argumentaban que se requería un gran número de pasos individualmente improbables para producir algo muy similar a un ser humano, un "humanoide"; que las probabilidades de que tal hecho ocurriera en otro planeta eran nulas; y que por lo tanto la probabilidad de la existencia de vida inteligente era cercana a cero. Pero claramente cuando hablamos de inteligencia extraterrestre, no estamos hablando (a pesar de *Star Trek*) de humanos o humanoides. Estamos hablando del equivalente funcional de los humanos –es decir, cualquier criatura capaz de construir y operar radiotelescopios. Podrían vivir en la tierra o en el mar o el aire. Podrían tener químicas, formas, tamaños, colores, extremidades y opiniones inimaginables. No estamos pensando que sigan la ruta particular que condujo a la evolución de los humanos. Debe haber muchos caminos evolutivos, cada uno improbable, pero la suma del número de caminos hacia la inteligencia puede ser, sin embargo, substancial.

En la presentación de Mayr, hay incluso un eco del artículo de Simpson (1964) *"La no prevalencia de humanoides"*. Pero el argumento básico es, creo, aceptable para todos nosotros. La evolución es oportunista y no pre-visiva. No "planea" desarrollar la inteligencia en unos miles de millones de años en el futuro. Responde a contingencias de corto plazo. Aun así, tal como sucede en otros campos, es mejor ser inteligente que ser estúpido, y se puede percibir una tendencia general hacia la inteligencia en los registros fósiles. En algunos mundos, la presión de la selección hacia la inteligencia puede ser más alta; en otros, más baja.

Si consideramos las estadísticas de nuestro propio caso –y tomamos un tiempo típico desde el origen de un sistema planetario al desarrollo de una civilización tecnológica de 4.600 millones de años– ¿qué sigue? No esperaríamos que las civilizaciones en mundos distintos evolucionaran en a un paso fijo. Algunas alcanzarían la inteligencia tecnológica más rápido, otras más lento, y "sin lugar a dudas" algunas nunca. Pero la Vía Láctea está atiborrada de estrellas de segunda y tercera generación (es decir, aquellas que poseen elementos pesados) tan antiguas como 10.000 millones de años.

Imaginemos dos curvas: La primera es la escala temporal probable hacia la evolución de la inteligencia tecnológica. Comienza en valores muy bajos; y en unos pocos miles de millones de años podría tener un valor significativo; en 5.000 millones de años, alcanza el 50 por ciento. La segunda curva representa las edades de las estrellas tipo solar, algunas de las cuales son muy jóvenes –están naciendo ahora mismo– mientras que otras tienen 10.000 millones de años de edad.

Si convolucionamos estas dos curvas, encontramos que hay una probabilidad de civilizaciones tecnológicas en planetas de estrellas de edades muy diferentes –no mucha en las demasiado jóvenes, pero más y más en las de mayor edad. El caso más probable es que el proyecto SETI detecte primero una civilización considerablemente más avanzada que la nuestra. Para cada una de esas civilizaciones, habrá habido decenas de miles de millones, o más, de especies adicionales. El número de sucesos improbables que haya tenido que ser concatenado para que esas especies pudieran evolucionar es enorme, y quizás existan miembros de cada una de esas especies que se enorgullecen de ser los únicos seres inteligentes de todo el universo.

## ¿Las civilizaciones necesariamente desarrollarían la tecnología para SETI?

Es perfectamente posible imaginar civilizaciones de poetas o (quizás) guerreros de la Edad de Bronce que nunca se tropezaron con las ecuaciones de James C. Maxwell y los receptores de radio. Pero serían sustituidas por la selección natural. La Tierra se encuentra rodeada de una población de asteroides y cometas que hacen que ocasionalmente el planeta se vea golpeado por uno lo suficientemente grande como para provocar daños substanciales. El episodio más famoso es el evento K-T (el impacto de un objeto cercano a la Tierra que ocurrió al final del período Cretácico y comienzos del Terciario), 65 millones de años atrás, el cual extinguió a los dinosaurios y muchas otras especies en la Tierra. Pero la probabilidad de que ocurra un impacto capaz de destruir una civilización es de uno en 2.000 en el próximo siglo.

Ya está claro que necesitamos elaborar medios de detección y seguimiento de objetos cercanos a la Tierra, así como los medios para su intercepción y destrucción. Si fallamos en esta tarea, simplemente seremos destruidos.

Las civilizaciones del Valle del Indo, Sumeria, Egipcia y Griega, así como otras, no tuvieron que enfrentar esta crisis porque no vivieron lo suficiente. Cualquier civilización duradera, terrestre o extraterrestre, debe haberse enfrentado con este peligro. Otros sistemas solares tendrán flujos de cometas y asteroides más o menos intensos, pero en casi todos los casos los peligros han de ser substanciales para la supervivencia de la especie.

La radiotelemetría, el monitoreo por radar de asteroides, y el concepto del espectro electromagnético es parte fundamental de cualquier tecnología temprana necesaria para lidiar con semejante amenaza. Por ende, cualquier civilización duradera se verá forzada, por selección natural, a desarrollar una tecnología apropiada similar a la utilizada en las observaciones de SETI. Por supuesto que no existe ninguna necesidad de tener órganos sensitivos que "vean" en la región de radio. Simplemente alcanza con la física.

Como la perturbación y la colisión en los cinturones de asteroides y cometas es perpetua, la amenaza de asteroides y cometas es igualmente perpetua, y no existe ningún momento en el que la tecnología pueda ser retirada.

A la vez, un programa de observación SETI representa una pequeña fracción del costo que significa lidiar con la amenaza de los cometas y asteroides. Por cierto, no es para nada cierto que SETI sea "muy limitado, alcanzando sólo una parte de nuestra galaxia." De existir transmisores lo suficientemente potentes, podríamos usar utilizar las actuales observaciones de los programas SETI para analizar la información de galaxias distantes. Como las transmisiones extraterrestres más probables de detectar serían las correspondientes a las civilizaciones más antiguas, podríamos esperar que estas sean también lo suficientemente potentes como para ser detectadas por nuestros actuales receptores y analizadores espectrales. Esta ha sido una de las estrategias aplicadas en los proyectos META (Mega Channel Extra Terrestrial Assay) de la Universidad de Harvard y del IAR en Argentina.

### ¿Es el programa SETI una fantasía de los físicos?

Mayr ha sugerido repetidamente que los proponentes de SETI son casi exclusivamente físicos y astrónomos, mientras que los biólogos tienen una visión más profunda y acertada. Como las tecnologías implicadas están relacionadas con las ciencias físicas, es razonable que los astrónomos, físicos e ingenieros desempeñen un papel principal en los programas de investigación SETI. Pero en 1982, cuando realicé una petición internacional publicada en la prestigiosa revista *Science* urgiendo la respetabilidad de SETI, no tuve dificultad alguna en lograr que la firmaran distinguidos biólogos y bioquímicos, incluidos David Baltimore, Melvin Calvin, Francis Crick, Manfred Eigen, Thomas Eisner, Stephen Jay Gould, Matthew Meselson, Linus Pauling, David Raup, y Edward O. Wilson. En mis tempranas especulaciones sobre estos asuntos, fui muy entusiasmado por mi mentor en biología, H.J. Muller, Premio Nobel de Genética. La petición publicada proponía que, en lugar de especular sobre el tema, refutáramos la hipótesis mediante las observaciones:

"*Es unánime nuestra convicción de que la única prueba significativa de la existencia de inteligencia extraterrestre es experimental. Ningún argumento a priori sobre la materia puede ser concluyente ni puede ser usado como sustituto de un programa observacional.*"

### Referencias

## 4.  Discusión

Ernst Mayr:

Entiendo completamente que la naturaleza de nuestro objeto de estudio nos permite solamente estimar probabilidades. No hay discrepancia alguna entre Carl Sagan y yo en cuanto a la existencia de grandes números de planetas en nuestra galaxia y en galaxias cercanas. El asunto, como fue enfatizado correctamente por Sagan, es la probabilidad de la evolución de una inteligencia superior y consecuentemente de una civilización tecnológica en un mundo habitado.

Una vez que tenemos vida (y casi seguramente será muy distinta de la vida en la Tierra), ¿cuál es la probabilidad de que desarrolle un linaje con inteligencia superior? En la Tierra, de los millones de linajes de organismos y quizás 50.000 millones de eventos de especialización, sólo uno condujo a la inteligencia superior; esto me hace creer en su absoluta improbabilidad.

Sagan adopta el principio "más listo es mejor" pero la vida en la Tierra refuta esta afirmación. Entre todas las formas de vida, ni los procariotas ni los protistas, hongos o plantas han necesitado desarrollar la inteligencia, como lo habrían hecho si fuese "mejor". En los más de 28 filos de animales, la inteligencia evolucionó sólo en uno (cordados) y con ciertas dudas también en los cefalópodos. Y en las miles de subdivisiones de los cordados, la inteligencia superior se desarrolló en un sólo caso, los primates, e incluso allí en una pequeña subdivisión. Esto no demuestra la supuesta inevitabilidad del desarrollo de la inteligencia superior porque "es mejor ser listo".

Sagan aplica un enfoque físico a este problema. Construye dos curvas lineales, ambas basadas en un pensamiento estrictamente determinista. Tal pensamiento es a menudo legítimo para los fenómenos físicos, pero es bastante inapropiado para eventos evolutivos o procesos sociales como el origen de las civilizaciones.

El argumento de que los extraterrestres, si perteneciesen a una civilización longeva, se verían forzados por la selección natural a desarrollar un saber tecnológico para enfrentar el peligro de los impactos cometarios es totalmente irrealista. ¿Cómo serían seleccionados los sobrevivientes de impactos previos para desarrollar el saber de la tecnología electrónica necesaria? Asimismo, el caso de la Tierra muestra cuán improbable es el origen de cualquier civiliza-ción a menos que la inteligencia superior se desarrolle primero. La Tierra asimismo muestra que las civilizaciones tienen inevitablemente corta duración.

Es una cuestión de sentido común que la existencia de inteligencia extraterrestre no puede ser establecida por argumentos *a priori*. Pero esto no justifica el proyecto SETI, ya que puede ser demostrado que el éxito de un programa observacional es tan improbable que puede, para fines prácticos, ser considerado cero.

En suma, no tengo la impresión de que los argumentos de Sagan hayan refutado los míos al grado de debilitarlos.

Carl Sagan

¿Es oportuno basarse en la vida en la Tierra? La esencia del argumento del Profesor Mayr es básicamente recorrer los varios factores de la ecuación de Drake (ver Shklovskii y Sagan, 1966) y asignarle valores cualitativos a cada uno. Él y yo estamos de acuerdo en que es esperable que las probabilidades de abundancia de planetas y orígenes de la vida sean altos. Enfatizo que los últimos resultados (Doyle 1995) sugieren uno o incluso dos planetas tipo Tierra con abundante agua líquida superficial, por cada sistema planetario en la galaxia.

Donde Mayr y yo discrepamos es en los últimos factores de la ecuación de Drake, especialmente en aquellos que atañen la probabilidad de la evolución de la inteligencia y las civilizaciones tecnológicas.

Mayr argumenta que los procariotas y protistas no han "evolucionado hacia la inteligencia". A pesar del gran respeto que le tengo al Profesor Mayr, debo objetar: los procariotas y los protistas son nuestros ancestros. Por lo tanto, han evolucionado hacia la inteligencia, junto con la mayoría del resto de la hermosa diversidad de la vida en la Tierra.

Por un lado, cuando destaca la pequeña fracción de especies que tienen inteligencia tecnológica, Mayr argumenta a favor de la relevancia de la vida en la Tierra en el problema de la inteligencia extraterrestre. Pero por otro lado, desprecia el ejemplo de la vida en la Tierra cuando ignora el hecho de que la inteligencia ha surgido aquí cuando nuestro planeta tiene otros cinco mil millones de años más de evolución por delante. Si fuera legítimo extrapolar desde el único ejemplo que tenemos frente a nosotros de la vida planetaria, se deduciría lo siguiente:

1. Hay enormes cantidades de planetas tipo Tierra, cada uno colmado de enormes cantidades de especies, y
2. En mucho menos del tiempo de evolución estelar en cada sistema planetario, al menos una de esas especies va a desarrollar inteligencia superior y tecnología.

Alternativamente, podríamos argumentar que es inapropiado extrapolar a partir de un sólo ejemplo. Pero entonces el argumento de uno en 50.000 millones de Mayr colapsa. Se me ocurre que él no puede plantearlo de la manera que le conviene según el caso. Con respecto a la evolución de la tecnología, cabe destacar que los chimpancés y los bonobos tienen cultura y tecnología. No sólo usan herramientas sino que también las elaboran a propósito para su uso futuro (ver Sagan and Druyan, 1992). De hecho, el bonobo Kanzi ha descubierto cómo manufacturar herramientas de piedra.

Es cierto, como destaca Mayr, que de las principales civilizaciones humanas, sólo una ha desarrollado la tecnología de la radiocomunicación. Pero esto dice muy poco acerca de la probabilidad que tiene una civilización humana de desarrollar tal tecnología. La misma civilización con radiotelescopios ha estado también a la vanguardia de la tecnología armamentista. Si, por ejemplo, la civilización de Europa occidental no hubiese destruido completamente a la civilización Azteca, ¿estos últimos habrían desarrollado radiotelescopios en el transcurso de siglos o milenios? Ya tenían un calendario astronómico superior al de los conquistadores.

Civilizaciones y especies apenas más capaces podrían tener la habilidad de eliminar la competencia. Pero esto no significa que la competencia no hubiese podido desarrollar capacidades comparables de no haber sido perturbados.

Mayr asegura que las plantas no reciben señales "electrónicas". Asumo que a lo que se refiere son señales "electromagnéticas". Pero las plantas lo hacen. Su existencia fundamental depende de recibir radiación electromagnética del Sol. La fotosíntesis y el fototropismo se pueden encontrar no sólo en las plantas más simples sino también en los protistas.

Todas las estrellas emiten luz visible, y las estrellas tipo solar emiten la mayoría de la radiación electromagnética en la parte visible del espectro. La percepción de la luz es una manera mucho más efectiva de comprender el ambiente a cierta distancia; ciertamente mucho más potente que los rastros olfativos. Es difícil imaginar una civilización tecnológica competente que no preste la debida atención a sus principales medios de examinar el mundo ex-terior. Incluso si usaran principalmente luz visible, ultravioleta, o infrarroja, la física es exactamente la misma que la de las ondas de radio; la diferencia es simplemente una cuestión de longitud de onda.

No insisto en que los argumentos anteriores sean convincentes, pero tampoco lo son los opuestos. No hemos podido investigar aún la evolución de biosferas en una gran gama de planetas. No hemos observado muchos casos de lo que es posible y lo que no. Hasta que no tengamos tal experiencia –o detectemos inteligencia extraterrestre- estaremos ciertamente envueltos en la incertidumbre.

La idea de que podamos, por medio de argumentos *a priori*, excluir la posibilidad de vida inteligente en los posibles planetas de las 400 mil millones de estrellas en la Vía Láctea me suena rara. Me recuerda a la gran cantidad de presunciones humanas que nos ubicaban en el centro del universo, o que nos diferenciaban no sólo en grado sino también en esencia del resto de la vida en la Tierra, o incluso afirmaban que el universo fue hecho en nuestro beneficio (Sagan, 1994). Comenzando con Copérnico, se ha mostrado que cada una de estas presunciones resulta carente de fundamentos.

En el caso de la inteligencia extraterrestre, admitamos nuestra ignorancia, pongamos a un lado los argumentos *a priori* y usemos la tecnología que tenemos la suerte de haber desarrollado para tratar de encontrar la respuesta. Esto es, pienso, lo que Charles Darwin –quien se convirtió de la religión ortodoxa a la biología evolutiva por el peso de evidencias observacionales– hubiera defendido.

Ernst Mayr

Como todo con lo que lidiamos son probabilidades, la mayoría de ellas extrapoladas de una muestra de uno, permítanme hacer unas pocas observaciones con respecto a la respuesta de Carl Sagan:

(1) No tenemos evidencia alguna del enorme número de planetas tipo Tierra "cada uno colmado de enormes cantidades de especies."

(2) Hay un mundo de diferencias entre plantas fotosintéticas y una civilización que desarrolla las teorías necesarias y la instrumentación para la comunicación electrónica.

(3) Sagan afirma "No hemos podido investigar aún la evolución de biosferas en una gran gama de planetas", la verdad es que no la hemos investigado en ningún planeta fuera de la Tierra.

(4) No estoy hablando de la posibilidad de vida inteligente; estoy hablando de la probabilidad de establecer un contacto con los medios disponibles. Ninguno de los argumentos de Sagan ha debilitado mi argumento de que ésta es prácticamente nula. Ésta no es una presunción, sino un cálculo serio de probabilidades. La respuesta negativa que seguramente tendremos no nos dirá nada acerca de la posibilidad de la inteligencia extraterrestre en algún lugar.

Carl Sagan

Saqué la conclusión tentativa de que otros planetas tipo terrestres tienen millones de especies de vida en ellos, del mismo conjunto de datos que conduce al Profesor Mayr a concluir que no hay civilizaciones técnicas extraterrestres. Mayr ahora concede (su observación 4) que puede haber inteligencia extraterrestre. ¿Quizás grandes números de planetas habitados por vida inteligente? Pero él duda acerca de si esta inteligencia puede haber desarrollado los medios necesarios para comunicaciones interestelares a través de ondas electromagnéticas. Como he enfatizado, no existe forma convincente alguna de evaluar este interrogante excepto, buscando transmisiones de radio interestelares. Eso es lo que estamos haciendo.